\documentclass[12pt]{article}

\usepackage{amsmath,amssymb,graphicx,color,array,cite}
\numberwithin{equation}{section}

\evensidemargin \oddsidemargin

\begin{document}

\newcommand{\story}{\vspace{5mm} \noindent $\spadesuit$ }

\begin{titlepage}

\renewcommand{\thefootnote}{\fnsymbol{footnote}}


\begin{flushright}
CQUeST-2008-0218
\end{flushright}

\vspace{15mm}
\baselineskip 9mm
\begin{center}
  {\Large \bf 
  Smearing Effect in Plane-Wave Matrix Model
  }
\end{center}

\baselineskip 6mm
\vspace{25mm}
\begin{center}
  Hyeonjoon Shin\footnote{\tt hshin@sogang.ac.kr} 
  \\[3mm] 
  {\sl Center for Quantum Spacetime\\ 
    Sogang University, Seoul
    121-742, South Korea }
  \\[3mm]
\end{center}

\thispagestyle{empty}

\vfill
\begin{center}
{\bf Abstract}
\end{center}
\noindent
Motivated by the usual D2-D0 system, we consider a configuration
composed of flat membrane and fuzzy sphere membrane in plane-wave
matrix model, and investigate the interaction between them.  The
configuration is shown to lead to a non-trivial interaction potential,
which indicates that the fuzzy sphere membrane really behaves like a
graviton, giant graviton.  Interestingly, the interaction is of
$r^{-3}$ type rather than $r^{-5}$ type.  We interpret it as the
interaction incorporating the smearing effect due to the fact that the
considered supersymmetric flat membrane should span and spin in four
dimensional subspace of plane-wave geometry.
\\ [5mm]

\vspace{5mm}
\end{titlepage}

\baselineskip 6.6mm
\renewcommand{\thefootnote}{\arabic{footnote}}
\setcounter{footnote}{0}

\section{Introduction}

The plane-wave or BMN matrix model \cite{Berenstein:2002jq} has been 
given by a mass deformation of the matrix model in flat spacetime, the 
BFSS matrix model \cite{Banks:1997vh}, and preserves full eleven
dimensional supersymmetry.  It has been believed to describe the 
M-theory in maximally supersymmetric plane-wave background in the 
framework of the discrete light cone quantization (DLCQ).

One peculiar property of the plane-wave matrix model which has
attracted much attention is that the supersymmetric fuzzy sphere
membrane with finite size appears from the vacuum structure in this
model \cite{Berenstein:2002jq,Dasgupta:2002hx}.  Although it is a
configuration of membrane, it has been interpreted as a graviton, or
more precisely a giant graviton because it has a size.  After finding
the presence of the fuzzy sphere membrane, there have been lots of
work studying its properties and related issues from various
viewpoints \cite{Dasgupta:2002hx}-\cite{Bak:2005jh}.  In the study of
dynamical aspect, it has been shown that the fuzzy sphere behaves
indeed like a graviton, and evidences about its interpretation as a
giant graviton have been accumulated
\cite{Lee:2003kf}-\cite{Yoshida:2005de}.  The thermodynamical aspect
has also been considered, and the vacuum structure involving fuzzy
sphere membranes at finite temperature has been investigated
\cite{Shin:2004ms}-\cite{Kawahara:2007nw}.  Upon a proper circle
compactification, the plane-wave matrix model leads to the matrix
string theory, which is related in the infrared limit to the free
string theory in ten-dimensional plane wave background
\cite{Sugiyama:2002tf}-\cite{Lozano:2006jr}.  This string theory
contains fuzzy spheres in its spectrum, whose various aspects also
have been studied in Refs.~\cite{Hyun:2003ks}-\cite{Das:2005vd}.

As for the dynamics of fuzzy sphere membrane, the research has been
focused on the interaction between the fuzzy sphere membranes
themselves.  Interaction between different kinds of membranes or other
supersymmetric objects in the plane-wave matrix model has not been
considered seriously.  In this paper, we are interested in the
configuration composed of the fuzzy sphere and flat membranes, each of
which is supersymmetric object, and investigate the interaction
between them.  If the interpretation of the fuzzy sphere membrane is
definitely correct, the configuration may be thought to have a
similarity with the usual D2-D0 system or more directly the
membrane-graviton system \cite{Aharony:1996bh} from eleven dimensional
point of view.  From this similarity, we may expect that the
interaction computed from the path integration of the plane-wave
matrix model around our configuration is the same up to numerical
constant with that in the D2-D0 system.  This expectation based on the
well-known D2-D0 system motivates the present study.  However, as we
will show, our expectation is partially correct.  The resulting
interacting potential at large $r$ distance gives one more evidence
that the fuzzy sphere membrane behaves like a graviton, that is, a
giant graviton, but has the $r^{-3}$ type rather than the expected
$r^{-5}$ type.  We will give an interpretation that the potential
incorporates the delocalization or smearing effect due to the
configuration of the supersymmetric flat membrane which should span
and spin in four dimensional space.

The organization of this paper is as follows.  In the next section, we
will give an expansion of the plane-wave matrix model around a general
classical background.  The background configuration composed of the
fuzzy sphere and flat membranes is presented in Sec.~\ref{bg}.  In
Sec.~\ref{1-loop-path-int}, the formal one-loop path integration of
the plane-wave matrix model around the background configuration of
Sec.~\ref{bg} is performed.  From the result of path integration, the
one-loop effective potential is obtained in Sec.~\ref{epotential}.
Finally, we give the conclusion and discussion in Sec.~\ref{finalsec}.

\section{Plane-wave matrix model}

The plane-wave or BMN matrix model \cite{Berenstein:2002jq} is a model
for the microscopic description of the DLCQ M-theory in the
eleven-dimensional $pp$-wave or plane-wave background
\cite{Kowalski-Glikman:1984wv}, which is $SO(3) \times SO(6)$
symmetric and given by
\begin{align}
ds^2 &= -2 dx^+ dx^- 
    - \left( \sum^3_{i=1} \left( \frac{\mu}{3} \right)^2 (x^i)^2
            +\sum^9_{a=4} \left( \frac{\mu}{6} \right)^2 (x^a)^2
      \right) (dx^+)^2
    + \sum^9_{I=1} (dx^I)^2 ~,
     \notag \\
F_{+123} &= \mu ~,
\label{pp}
\end{align}
with the index notation $I=(i,a)$. This background is maximally
supersymmetric and obtained by taking the Penrose limit to the
eleven-dimensional AdS type geometries \cite{Blau:2002dy}.

The plane-wave matrix model is basically composed of two parts.  One
part is the usual matrix model based on eleven-dimensional flat
space-time, that is, the flat space matrix model, and another is a set
of terms reflecting the structure of the maximally supersymmetric
eleven dimensional plane-wave background, Eq. (\ref{pp}).  Its action
is
\begin{equation}
S_{pp} = S_\mathrm{flat} + S_\mu ~,
\label{pp-bmn}
\end{equation}
where each part of the action on the right hand side is given by
\begin{align}
S_\mathrm{flat} & = \int dt \mathrm{Tr} 
\left( \frac{1}{2R} D_t X^I D_t X^I + \frac{R}{4} ( [ X^I, X^J] )^2
      + i \Theta^\dagger D_t \Theta 
      - R \Theta^\dagger \gamma^I [ \Theta, X^I ]
\right) ~,
  \notag \\
S_\mu &= \int dt \mathrm{Tr}
\left( 
      -\frac{1}{2R} \left( \frac{\mu}{3} \right)^2 (X^i)^2
      -\frac{1}{2R} \left( \frac{\mu}{6} \right)^2 (X^a)^2
      - i \frac{\mu}{3} \epsilon^{ijk} X^i X^j X^k
      - i \frac{\mu}{4} \Theta^\dagger \gamma^{123} \Theta
\right) \, .
\label{o-action}
\end{align}
Here, $R$ is the radius of circle compactification along $x^-$, $D_t$
is the covariant derivative with the gauge field $A$,
\begin{equation}
D_t = \partial_t - i [A, \: ] ~,
\end{equation}
and $\gamma^I$ is the $16 \times 16$ $SO(9)$ gamma matrices.  For
practical study of the model, it is often convenient to make $R$
disappear from the action by taking the rescaling of the gauge field
and parameters as
\begin{equation}
A \rightarrow R A \, ,~~~ 
t \rightarrow \frac{1}{R} t \, ,~~~
\mu \rightarrow R \mu \, .
\end{equation}
Then the actions in Eq.~(\ref{o-action}) become
\begin{align}
S_\mathrm{flat} & = \int dt \mathrm{Tr} 
\left( \frac{1}{2} D_t X^I D_t X^I + \frac{1}{4} ( [ X^I, X^J] )^2
      + i \Theta^\dagger D_t \Theta 
      -  \Theta^\dagger \gamma^I [ \Theta, X^I ]
\right) \, ,
  \notag \\
S_\mu &= \int dt \mathrm{Tr}
\left( 
      -\frac{1}{2} \left( \frac{\mu}{3} \right)^2 (X^i)^2
      -\frac{1}{2} \left( \frac{\mu}{6} \right)^2 (X^a)^2
      - i \frac{\mu}{3} \epsilon^{ijk} X^i X^j X^k
      - i \frac{\mu}{4} \Theta^\dagger \gamma^{123} \Theta
\right) \, ,
\label{pp-action}
\end{align}
which are free of $R$.

In matrix model, various objects, like branes and graviton, are
realized by the classical solutions of the equations of motion for the
matrix field.  The dynamics between them is studied by expanding the
matrix model action around the corresponding classical solution and
performing the path integration.  Let us denote the classical solution
or the background configuration by $B^I$, and split the matrix
quantities into as follows:
\begin{equation}
X^I = B^I + Y^I \, , \quad A=0+A \, , \quad \Theta = 0 + \Psi \, .
\label{cl+qu}
\end{equation}
Then $Y^I$, $A$ and $\Psi$ are the quantum fluctuations around the
background configuration, which are the fields subject to the path
integration.  We note that the gauge field may also have non-trivial
classical configuration.  However, it is simply set to zero in this
paper because the objects we are interested in do not generate any
background gauge field.

In taking into account the quantum fluctuations, we should recall that
the matrix model itself is a gauge theory.  This implies that the
gauge fixing condition should be specified before proceed further.  In
this paper, we take the background field gauge which is usually chosen
in the matrix model calculation,
\begin{equation}
D_\mu^{\rm bg} A^\mu_{\rm qu} \equiv
D_t A + i [ B^I, X^I ] = 0 ~.
\label{bg-gauge}
\end{equation}
Then the corresponding gauge-fixing $S_\mathrm{GF}$ and Faddeev-Popov
ghost $S_\mathrm{FP}$ terms are given by
\begin{equation}
S_\mathrm{GF} + S_\mathrm{FP} =  \int\!dt \,{\rm Tr}
  \left(
      -  \frac{1}{2} (D_\mu^{\rm bg} A^\mu_{\rm qu} )^2 
      -  \bar{C} \partial_{t} D_t C + [B^I, \bar{C}] [X^I,\,C]
\right) ~.
\label{gf-fp}
\end{equation}

Now by inserting the decomposition of the matrix fields (\ref{cl+qu})
into Eqs.~(\ref{pp-action}) and (\ref{gf-fp}), we get the gauge fixed
plane-wave action $S$ $(\equiv S_{pp} + S_\mathrm{GF} +
S_\mathrm{FP})$ expanded around the classical background $B^I$.  The
resulting action is read as
\begin{equation}
S =  S_0 + S_2 + S_3 + S_4 ~,
\end{equation}
where $S_n$ represents the action of order $n$ with respect to the
quantum fluctuations and, for each $n$, its expression is
\begin{align}
S_0 = \int dt \, \mathrm{Tr} \bigg[ \,
&      \frac{1}{2}(\dot{B}^I)^2  
        - \frac{1}{2} \left(\frac{\mu}{3}\right)^2 (B^i)^2 
        - \frac{1}{2} \left(\frac{\mu}{6}\right)^2 (B^a)^2 
        + \frac{1}{4}([B^I,\,B^J])^2
        - i \frac{\mu}{3} \epsilon^{ijk} B^i B^j B^k 
    \bigg] ~,
\notag \\
S_2 = \int dt \, \mathrm{Tr} \bigg[ \,
&       \frac{1}{2} ( \dot{Y}^I)^2 - 2i \dot{B}^I [A, \, Y^I] 
        + \frac{1}{2}([B^I , \, Y^J])^2 
        + [B^I , \, B^J] [Y^I , \, Y^J]
        - i \mu \epsilon^{ijk} B^i Y^j Y^k
\notag \\
&       - \frac{1}{2} \left( \frac{\mu}{3} \right)^2 (Y^i)^2 
        - \frac{1}{2} \left( \frac{\mu}{6} \right)^2 (Y^a)^2 
        + i \Psi^\dagger \dot{\Psi} 
        -  \Psi^\dagger \gamma^I [ \Psi , \, B^I ] 
        -i \frac{\mu}{4} \Psi^\dagger \gamma^{123} \Psi  
\notag \\ 
&       - \frac{1}{2} \dot{A}^2  - \frac{1}{2} ( [B^I , \, A])^2 
        + \dot{\bar{C}} \dot{C} 
        + [B^I , \, \bar{C} ] [ B^I ,\, C] \,
     \bigg] ~,
\notag \\
S_3 = \int dt \, \mathrm{Tr} \bigg[
&       - i\dot{Y}^I [ A , \, Y^I ] - [A , \, B^I] [ A, \, Y^I] 
        + [ B^I , \, Y^J] [Y^I , \, Y^J] 
        +  \Psi^\dagger [A , \, \Psi] 
\notag \\
&       -  \Psi^\dagger \gamma^I [ \Psi , \, Y^I ] 
        - i \frac{\mu}{3} \epsilon^{ijk} Y^i Y^j Y^k
        - i \dot{\bar{C}} [A , \, C] 
        +  [B^I,\, \bar{C} ] [Y^I,\,C]  \,
     \bigg] ~,
\notag \\
S_4 = \int dt \, \mathrm{Tr} \bigg[
&       - \frac{1}{2} ([A,\,Y^I])^2 + \frac{1}{4} ([Y^I,\,Y^J])^2 
     \bigg] ~.
\label{bgaction} 
\end{align}

\section{Background configuration}
\label{bg}

In this section, we set up the background configuration corresponding
to the flat membrane and fuzzy sphere membrane, and discuss about the
perturbation theory around it.

Since we will study the interaction between two objects, the matrices
representing the background have the $2 \times 2$ block diagonal form
as
\begin{equation}
B^I = \begin{pmatrix} B_{(1)}^I & 0 \\ 0 & B_{(2)}^I
      \end{pmatrix} \, ,
\label{gb}
\end{equation}
where $B_{(s)}^I$ with $s=1,2$ are $N_s \times N_s$ matrices.  If
$B^I$ are taken to be $N \times N$ matrices, then $N = N_1 + N_2$.

Basically, the configuration we consider is that the fuzzy sphere
membrane is placed in the transverse space of the flat membrane with
distance $r$.  Each membrane is supposed to be supersymmetric.  We
would like to note that, unlike the case of membrane placed in flat
space-time, supersymmetric membrane in plane-wave background may have
a particular motion in a given situation.  This feature stems from the
nature of the plane-wave background.

The first object corresponding to $B^I_{(1)}$ is taken to be the fuzzy
sphere membrane, which spans in $SO(3)$ symmetric space and rotates in
$x^8$-$x^9$ plane as follows:
\begin{gather}
B^i_{(1)} = \frac{\mu}{3} J^i ~, \notag \\
B^8_{(1)} = r \cos(\mu t/6) \mathbf{1}_{N_1 \times N_1}~, 
 \quad B^9_{(1)} = r \sin(\mu t/6) \mathbf{1}_{N_1 \times N_1}
\label{fuzzy}
\end{gather}
where $J^i$ is in the $N_1$-dimensional irreducible representation of
$SU(2)$ and thus satisfies the $SU(2)$ algebra,
\begin{equation}
[ J^i, J^j ] = i \epsilon^{ijk} J^k ~.
\label{su2}
\end{equation}
If the fuzzy sphere membrane sits at the origin in the $SO(6)$
symmetric space, it preserves the full 16 dynamical supersymmetries of
the plane-wave and hence is 1/2-BPS object.  The above configuration
contains a circular motion, and thus seems to break the supersymmetry.
However, as has been shown explicitly in the path integral formulation
\cite{Shin:2003np}, it is still supersymmetric basically due to the
presence of the plane-wave background as alluded to above.  In
addition to this, the value of the classical action simply vanishes
without any velocity dependent term.  In this sense, the fuzzy sphere
in circular motion may be regarded as a `static' object.

The second object represented by $B^I_{(2)}$ is the flat membrane,
which is taken to be the one found in \cite{Hyun:2002cm}.  It is
1/8-BPS object, and spans and spins in four dimensional subspace of
the $SO(6)$ symmetric space as
\begin{gather}
B^4_{(2)} = Q \cos(\mu t/6) \, , \quad B^6_{(2)} = Q \sin(\mu t/6) \, ,
\notag \\
B^5_{(2)} = P \cos(\mu t/6) \, , \quad B^7_{(2)} = P \sin(\mu t/6) \, ,
\label{flat}
\end{gather}
where $N_2 \times N_2$ matrices, $Q$ and $P$, satisfy
\begin{align}
[ Q, P ] = i \sigma \, ,
\label{flatcom}
\end{align}
with a small constant parameter $\sigma$.  We note that, in order to
describe the flat membrane properly, the size of the matrix should be
infinite.  In what follows, $N_2$ is thus implicitly taken to be
infinite.  Now, from this somewhat complicated configuration, we see
that, at $t=0$, the flat membrane is placed in $x^4$-$x^5$ plane, and,
as time goes by, one axis along $x^4$ rotates in $x^4$-$x^6$ plane
while another axis along $x^5$ rotates in $x^5$-$x^7$ plane.

In spite of the circular motion of the fuzzy sphere membrane and the
spinning motion of the flat membrane, the configuration of
Eq.~(\ref{gb}) with Eqs.~(\ref{fuzzy}) and (\ref{flat}) is similar to
the configuration of D2 and D0 branes separated by a constant
distance, because the distance between flat and fuzzy sphere
membranes, $r$, does not change in time.

Having the background configuration (\ref{gb}), we first evaluate the
classical value of the action $S_0$.  Because all the motions involved
in the background have the same period, $T=12 \pi/\mu$, it is
sufficient to consider the action per one period, which is obtained as
\begin{equation}
S_0/T = - \frac{1}{2} N_2 \sigma^2 \, .
\label{clav}
\end{equation}

The next thing we are going to do in what follows is the computation
of the one-loop correction to this action, that is, to the background,
(\ref{fuzzy}) and (\ref{flat}), due to the quantum fluctuations via
the path integration of the quadratic action $S_2$, and obtain the
one-loop effective action $\Gamma_\mathrm{eff}$ or the effective
potential $V_\mathrm{eff}$.  Before doing the one-loop computation, it
should be made clear that $S_3$ and $S_4$ of Eq.~(\ref{bgaction}) can
be regarded as perturbations.  For this purpose, following
\cite{Dasgupta:2002hx}, we rescale the fluctuations and parameters as
\begin{gather}
A   \rightarrow \mu^{-1/2} A   \, , \quad
Y^I \rightarrow \mu^{-1/2} Y^I \, , \quad
C   \rightarrow \mu^{-1/2} C   \, , \quad
\bar{C} \rightarrow \mu^{-1/2} \bar{C} \, ,
\notag \\
r \rightarrow \mu r \, , \quad
t \rightarrow \mu^{-1} t \, , \quad
Q \rightarrow \mu Q \, , \quad
P \rightarrow \mu P \, , \quad
\sigma \rightarrow \mu^2 \sigma \, .
\label{rescale}
\end{gather}
Under this rescaling, the powers of $\mu$ are factored out from the
action $S$ in the background (\ref{fuzzy}) and (\ref{flat}) as
\begin{align}
S =  \mu^3 S_0 + S_2 + \mu^{-3/2} S_3 + \mu^{-3} S_4 ~,
\label{ssss}
\end{align}
where $S_0$, $S_2$, $S_3$ and $S_4$ do not have $\mu$ dependence and
the period of motion becomes $12\pi$.  Now it is obvious that, in the
large $\mu$ limit, $S_3$ and $S_4$ can be treated as perturbations and
the one-loop computation gives the sensible result.

Based on the structure of (\ref{gb}), we now write the quantum
fluctuations in the $2 \times 2$ block matrix form as follows:
\begin{gather}
A   = \begin{pmatrix}
          0               &   \Phi^0      \\
         \Phi^{0 \dagger} &   0
      \end{pmatrix} ~,~~~
Y^I = \begin{pmatrix}
          0               &   \Phi^I      \\
         \Phi^{I \dagger} &   0
      \end{pmatrix} ~,~~~
\Psi = \begin{pmatrix}
          0             &  \chi \\
         \chi^{\dagger} &  0
       \end{pmatrix} ~,
 \notag \\
C =    \begin{pmatrix}
         0           &  C  \\ 
         C^{\dagger} &  0
       \end{pmatrix} ~,~~~
\bar{C} = \begin{pmatrix}
              0               &  \bar{C}  \\
              \bar{C}^\dagger &  0
           \end{pmatrix} ~.
\label{q-fluct}
\end{gather}
Although we denote the block off-diagonal matrices for the ghosts by
the same symbols with those of the original ghost matrices, there will
be no confusion since $N \times N$ matrices will never appear from now
on.  The reason why the block-diagonal parts are not considered is
that they do not give any effect on the interaction between two kinds
of membranes but lead to the quantum correction to each membrane
itself, which vanishes because the membranes considered here are
supersymmetric.

\section{One-loop path integration}
\label{1-loop-path-int}

We now perform the path integration for the action of quadratic
fluctuations, $S_2$, around the classical background (\ref{gb}) with
(\ref{fuzzy}) and (\ref{flat}).  The results will be formal and the
actual evaluation of them for the effective action or potential will
be described in the next section.

The quadratic action is largely composed of three decoupled sectors,
which are bosonic, ghost, and fermionic sectors.  In the path
integration of each sector, the integration variables are matrices.
For the actual evaluation of the path integration, it is usually
useful to expand the matrix variables in a suitable matrix basis.
Taking a matrix basis depends on the classical background under
consideration.  For example, in the study of fuzzy spheres, the matrix
spherical harmonics provides a good matrix basis for the fluctuations
around the configuration of fuzzy spheres. For the present case where
the flat membrane is involved, the matrix spherical harmonics is not
adequate, and we should look for another basis.

For taking a suitable matrix basis, we first consider the fluctuations
around the fuzzy sphere and flat membranes separately.  The fuzzy
sphere of Eq.~(\ref{fuzzy}) is described by $N_1$-dimensional or
spin-$j$ representation of $SU(2)$ with
\begin{align}
j = \frac{1}{2} ( N_1 -1) \, .
\end{align}
Thus, the fluctuations around the fuzzy sphere are naturally expressed
in terms of the states $|m \rangle$ in the spin-$j$ representation of
$SU(2)$, where $-j \le m \le j$.  $J^i$ describing fuzzy sphere acts
on $|m \rangle$ in a standard way as
\begin{align}
J^3 |m\rangle = m |m\rangle \, , \quad
J^\pm |m\rangle = \sqrt{(j \mp m)(j \pm m +1)} |m\rangle \, ,
\label{su2alg}
\end{align}
where $J^\pm = J^1 \pm i J^2$.  As for the flat membrane of
Eq.~(\ref{flat}), it has the characteristic given by the commutation
relation, (\ref{flatcom}).  If we define
\begin{align}
a = \frac{1}{\sqrt{2 \sigma}} (Q+iP) ~, \quad
a^\dagger = \frac{1}{\sqrt{2 \sigma}} (Q-iP)~,
\label{osc-ca}
\end{align}
then they satisfy the commutation relation
\begin{align}
[ a, a^\dagger ] =1 \, ,
\end{align}
and can be regarded as the annihilation and creation operators of
simple harmonic oscillator. This fact allows us to express the
fluctuations around the flat membrane in terms of the oscillator
states, on which $a$ and $a^\dagger$ act as
\begin{align}
a | n \rangle = \sqrt{n} | n-1 \rangle \, , \quad
a^\dagger | n \rangle = \sqrt{n+1} | n+1 \rangle \, .
\label{osc-alg}
\end{align}
Because the size of the membrane is given by $N_2$, the oscillator
number $n$ runs from $0$ to $N_2-1$, and hence has the upper bound.
However, we note that actually there is no upper bound for $n$ because
$N_2$ should be infinite for the proper description of the flat
membrane.

From the above consideration and the structure of Eq.~(\ref{q-fluct}),
the matrix basis for the fluctuation can be taken to be $| m \rangle
\langle n|$, where $|m \rangle$ is the state in spin-$j$
representation of $SU(2)$ and $\langle n|$ is an oscillator state.
Then, in this matrix basis, each fluctuation matrix has the following
mode expansion
\begin{align}
\Phi = \sum_{m=-j}^j \sum_{n=0}^{\infty} \phi_{mn} 
| m \rangle \langle n | \, .
\label{decom}
\end{align}
This expansion now allows us to reduces the path integration of the
matrix variable to that of the mode $\phi_{mn}$.

\subsection{Bosonic sector}
 
The Lagrangian for the bosonic sector of the quadratic action $S_2$ is
split into two parts
\begin{align}
L_B = L_{SO(3)} + L_{SO(6)} \, ,
\end{align}
where $L_{SO(3)}$ is the Lagrangian for $\Phi^i$ and $L_{SO(6)}$ is
for $\Phi^0$ and $\Phi^a$.  Because two parts are decoupled systems,
each of them can be considered independently.

We fist deal with the path integration of $L_{SO(3)}$.  
The Lagrangian is
\begin{align}
L_{SO(3)} = \mathrm{Tr}
 \left[ \,
  | \dot{\Phi}^i |^2 - (r^2 +Q^2 + P^2) | \Phi^i |^2
  - \frac{1}{3^2} | \Phi^i + i \epsilon^{ijk} J^j \Phi^k |^2
  - \frac{1}{3^2} \Phi^{i \dagger} J^i J^j \Phi^j
 \, \right] \, .
\end{align}
Due to the third term in the trace, the diagonalization of $\Phi^i$ is
required.  The procedure of diagonalization has been well established
based on the mode expansion (\ref{decom}) and the standard $SU(2)$
algebra \cite{Dasgupta:2002hx,Shin:2003np}.  If we adopt the procedure
with the same symbols used in previous literatures, the
diagonalization of the modes of $\Phi^i$, that is, $\phi^i_{mn}$,
leads to $\alpha_{mn}$, $\beta_{mn}$, and $\omega_{mn}$, which are
described by the following Lagrangian.
\begin{align}
L_{SO(3)} =
&  \sum^{j-1}_{m=-j+1} \sum^\infty_{n=0}
\left[  |\dot{\alpha}_{mn}|^2 
  - \left( r^2 + \sigma (2n+1)+\frac{1}{3^2} j^2 \right)
    | \alpha_{mn} |^2 
\right]
  \notag \\
&  + \sum^{j+1}_{m=-j-1} \sum^\infty_{n=0}
\left[  | \dot{\beta}_{mn} |^2
  - \left( r^2 + \sigma (2n+1)+ \frac{1}{3^2} (j+1)^2 \right) 
    | \beta_{mn} |^2 
\right]
  \notag \\
&  + \sum^{j}_{m=-j} \sum^\infty_{n=0}
\left[
 | \dot{\omega}_{mn} |^2 
- \left( r^2 + \sigma (2n+1) + \frac{1}{3^2} j (j+1) \right)
   | \omega_{mn} |^2
\right] \, ,
\end{align}
where the range of $m$ has been changed due to the effect of 
diagonalization and 
\begin{align}
(Q^2+P^2) | n \rangle = \sigma (2 a^\dagger a +1 ) |n \rangle
=\sigma (2 n +1 ) | n \rangle
\end{align}
has been used.  As noted in \cite{Dasgupta:2002hx,Shin:2003np}, the
mode $\omega_{mn}$ corresponds to the gauge degree of freedom and its
effect should be cancelled by the contribution from ghosts.  Now,
having the fully diagonalized Lagrangian, it is straightforward to
perform the path integration and get
\begin{align}
\prod^\infty_{n=0}  
\det{}^{-(2j-1)} \Delta_{(n)10}  
\cdot \det{}^{-(2j+3)} \Delta_{(n)11}   
\cdot \det{}^{-(2j+1)} \Delta_{(n)} \, .
\label{so3}
\end{align}
where we have defined
\begin{align}
\Delta_{(n) \alpha \beta} \equiv 
   - \partial_t^2 - r^2 - \sigma (2n+\alpha) 
   - \frac{1}{3^2} (j+\beta)^2 \, ,
\notag \\
\Delta_{(n)} \equiv - \partial_t^2 - r^2 - \sigma (2n+1)
- \frac{1}{3^2} j(j+1) \, .
\label{kinop}
\end{align}

We turn to another part of the bosonic sector, which
is given by
\begin{align}
L_{SO(6)} = \mathrm{Tr}  
\bigg\{ 
& - | \dot{\Phi}^0 |^2 
  + (r^2 +Q^2 + P^2) | \Phi^0 |^2
  + \frac{1}{3^2} \Phi^{0 \dagger} J^i J^i  \Phi^0 
\notag \\
& + | \dot{\Phi}^a |^2 
  - \Big( r^2 + Q^2+P^2 +\frac{1}{6^2} \Big) | \Phi^a |^2
  - \frac{1}{3^2} \Phi^{a \dagger} J^i  J^i  \Phi^a 
\notag \\
& + \frac{i}{3} \sin (t/6)
  \left[
   \Phi^{0 \dagger} (\Phi^4 Q + \Phi^5 P - r \Phi^8 )   
    - (Q \Phi^{4\dagger} + P \Phi^{5 \dagger} 
    - r \Phi^{8 \dagger} ) \Phi^0 
  \right]
\notag \\
&  - \frac{i}{3} \cos ( t/6 )
  \left[
   \Phi^{0 \dagger} (\Phi^6 Q + \Phi^7 P - r \Phi^9 )   
    - (Q \Phi^{6\dagger} + P \Phi^{7 \dagger} 
    - r \Phi^{9 \dagger} ) \Phi^0 
  \right]  
\notag \\
&  + 2 i \sigma  
  \left[
    \left(\cos (t/6 )\Phi^{4 \dagger} 
      + \sin ( t/6 ) \Phi^{6 \dagger}
    \right)
    \left(\cos ( t/6 ) \Phi^5 
      + \sin ( t/6 ) \Phi^7
    \right)
  \right.
\notag \\
& \left.
   - \left( \cos ( t/6 ) \Phi^{5 \dagger} 
       + \sin ( t/6 ) \Phi^{7 \dagger}
    \right) 
    \left( \cos ( t/6 )\Phi^4 
         + \sin ( t/6 ) \Phi^6
    \right)
  \right]   
\bigg\} ~.
\label{lb}
\end{align}
There are lots of trigonometric functions in this Lagrangian due to
the motion of background membranes.  The fact that they have explicit
time dependence makes the path integration cumbersome.  Thus, it is
desirable to hide the explicit time dependence by taking some
redefinition of matrix variables.  For the present case, we take
\begin{align}
\cos(t/6) \Phi^4 + \sin(t/6) \Phi^6 
\rightarrow \Phi^4 \, ,\quad
-\sin(t/6) \Phi^4 + \cos(t/6) \Phi^6 
\rightarrow \Phi^6 \, ,
\notag \\
\cos(t/6) \Phi^5 + \sin(t/6) \Phi^7
\rightarrow \Phi^5 \, ,\quad
-\sin(t/6) \Phi^5 + \cos(t/6) \Phi^7
\rightarrow \Phi^7 \, ,
\notag \\
\cos(t/6) \Phi^8 + \sin(t/6) \Phi^9
\rightarrow \Phi^8 \, ,\quad
-\sin(t/6) \Phi^8 + \cos(t/6) \Phi^9
\rightarrow \Phi^9 \, ,
\end{align}
which is nothing but the transformation to the rotating frame.  Then,
under this transformation, the Lagrangian becomes
\begin{align}
L_{SO(6)} =\mathrm{Tr} \bigg\{ \,
& - | \dot{\Phi}^0 |^2 
  + (r^2 +Q^2 + P^2) | \Phi^0 |^2
  + \frac{1}{3^2} \Phi^{0 \dagger} J^i J^i  \Phi^0
\notag \\
& + | \dot{\Phi}^a |^2 
  - (r^2 +Q^2+P^2) | \Phi^a |^2 
  - \frac{1}{3^2} 
           \Phi^{a \dagger} J^i  J^i \Phi^a
 \notag \\
& +\frac{1}{3}  
     (\Phi^{4 \dagger} \dot{\Phi}^6
      - \Phi^{6 \dagger} \dot{\Phi}^4 ) 
  +\frac{1}{3} 
     (\Phi^{5 \dagger} \dot{\Phi}^7
      - \Phi^{7\dagger} \dot{\Phi}^5 ) 
  +\frac{1}{3}  
     (\Phi^{8 \dagger} \dot{\Phi}^9
      - \Phi^{9 \dagger} \dot{\Phi}^8 )         
\notag \\
& - \frac{i}{3} 
   \Phi^{0 \dagger} (\Phi^6 Q + \Phi^7 P - r \Phi^9 )   
  + \frac{i}{3} 
   (Q \Phi^{6 \dagger} + P \Phi^{7 \dagger} 
    - r \Phi^{9 \dagger} ) \Phi^0 
 \notag \\ 
& + 2 i \sigma (\Phi^{4 \dagger} \Phi^5
      - \Phi^{5\dagger} \Phi^4 )   
 \, \bigg\} \, ,
\end{align}
which is obviously free of trigonometric functions having explicit
time dependence.

Now, by using the mode expansion Eq.~(\ref{decom}) for each matrix
variable, we can express this Lagrangian in terms of modes.  We notice
however that the terms linear in $Q$ and $P$ lead to coupling of modes
with different oscillator number $n$ because $Q$ and $P$ are linear
combinations of the creation and annihilation operators as seen in
Eq.~(\ref{osc-ca}).  In order to avoid such coupling, we follow the
prescription given in \cite{Aharony:1996bh} and define new matrix
variables as
\begin{align}
\Phi^\pm \equiv 
   \frac{1}{\sqrt{2}} ( \Phi^4 \pm i \Phi^5 )  \, ,
\quad
\tilde{\Phi}^\pm \equiv 
   \frac{1}{\sqrt{2}} ( \Phi^6 \pm i \Phi^7 ) \, ,
\end{align}
which is nothing but a unitary transformation.  Then the terms
linear in $Q$ and $P$ become
\begin{align}
& \frac{i}{3} \mathrm{Tr} \left\{ 
-  \Phi^{0 \dagger} (\Phi^6 Q + \Phi^7 P  )   
+  (Q \Phi^{6 \dagger} + P \Phi^{7 \dagger} ) \Phi^0
\right\}
\notag \\
=& \frac{i}{3} \sqrt{\sigma} \mathrm{Tr} \left\{
- \Phi^{0 \dagger} (\tilde{\Phi}^+ a^\dagger + \tilde{\Phi}^- a )   
+ (a \tilde{\Phi}^{+ \dagger} + a^\dagger \tilde{\Phi}^{- \dagger} ) 
  \Phi^0 \right\} \, ,
\end{align}
where Eq.~(\ref{osc-ca}) has been used.  This structure naturally
leads us to take the mode expansions for $\Phi^\pm$ and
$\tilde{\Phi}^\pm$ as
\begin{align}
\Phi^\pm = \sum_{m=-j}^j \sum_{n=\mp 1}^{\infty} \phi^\pm_{mn} 
| m \rangle \langle n \pm 1 | \, , \quad
\tilde{\Phi}^\pm = \sum_{m=-j}^j \sum_{n=\mp 1}^{\infty} 
\tilde{\phi}^\pm_{mn} | m \rangle \langle n \pm 1 | \, ,
\label{mode}
\end{align}
while $\Phi^0$, $\Phi^8$, and $\Phi^9$ are taken to follow the
expansion of Eq.~(\ref{decom}).  We note that $\Phi^\pm$ and
$\tilde{\Phi}^\pm$ should have the same type of mode expansion, since
they couple to each other with one time derivative.

Having proper mode expansions for matrix variables, there is no longer
mode mixing between different $n$ or $m$, and thus the Lagrangian is
the sum of parts each of which is labeled by $m$ and $n$.  For a given
$m$ and $n$, after some manipulation with Eqs.~(\ref{su2alg}) and
(\ref{osc-alg}), the part of the Lagrangian, say $L_{(mn)}$, is
obtained as
\begin{align}
L_{(mn)}=V_{(mn)}^\dagger M_{(mn)} V_{(mn)} \, ,
\end{align}
where
$ V_{(mn)} = (
\phi^0_{mn} , \, \phi^+_{mn} , \, \tilde{\phi}^+_{mn} , \,
\phi^-_{mn} , \, \tilde{\phi}^-_{mn} , \, \phi^8_{mn} , \,
\phi^9_{mn} )^T $ and
\begin{align}
M_{(mn)} = 
\begin{pmatrix}
-\Delta_{(n)} & 0 & -\frac{i}{3} \sqrt{\sigma}\sqrt{n+1} & 0 
       & -\frac{i}{3} \sqrt{\sigma}\sqrt{n}  & 0 & \frac{i}{3}r \\
0& \Delta_{(n)} & \frac{1}{3} \partial_t 
       & 0 & 0 & 0 & 0 \\
\frac{i}{3} \sqrt{\sigma}\sqrt{n+1} & -\frac{1}{3} \partial_t 
       & \Delta_{(n)}-2\sigma & 0 & 0 & 0 & 0 \\
0 & 0 & 0 & \Delta_{(n)} & \frac{1}{3} \partial_t & 0 & 0 \\
\frac{i}{3} \sqrt{\sigma}\sqrt{n} & 0 & 0 
       & -\frac{1}{3} \partial_t & \Delta_{(n)}+2\sigma & 0 & 0 \\
0 & 0 & 0 & 0 & 0 & \Delta_{(n)} & \frac{1}{3} \partial_t \\
-\frac{i}{3}r & 0 & 0 & 0 & 0 & -\frac{1}{3} \partial_t & 
\Delta_{(n)}
\end{pmatrix} \, ,
\end{align}
where $\Delta_{(n)}$ has been defined in Eq.~(\ref{kinop}).  Before
summing up $L_{(mn)}$ for $m$ and $n$, we should notice that the
oscillator number $n$ of $\phi^+_{mn}$ and $\tilde{\phi}^+_{mn}$
starts from $-1$ while that of $\phi^-_{mn}$ and $\tilde{\phi}^-_{mn}$
starts from $+1$, as we can see from Eq.~(\ref{mode}).  It is easy to
see that the modes $\phi^+_{mn}$ and $\tilde{\phi}^+_{mn}$ at $n=-1$
are decoupled from other modes and form a subsystem, because all other
modes do not have such oscillator number.  As for the modes
$\phi^-_{mn}$ and $\tilde{\phi}^-_{mn}$, the absence of them at $n=0$
seems to require an independent treatment of $M_{(m0)}$.  However, let
us suppose that these modes were present at the beginning.  Then, the
structure of $M_{(m0)}$ shows that they are decoupled from other modes
and form a subsystem.  Furthermore, the subsystem is exactly the same
with that composed of $\phi^+_{mn}$ and $\tilde{\phi}^+_{mn}$ at
$n=-1$.  This indicates that the modes $\phi^+_{m-1}$ and
$\tilde{\phi}^+_{m-1}$ can be symbolically identified with
$\phi^-_{m0}$ and $\tilde{\phi}^-_{m0}$.  More precisely,
$\phi^+_{m-1} \rightarrow \tilde{\phi}^-_{m0}$ and
$\tilde{\phi}^+_{m-1} \rightarrow \phi^-_{m0}$, which can be inferred
from $M_{(m0)}$.  After all, all the modes can be taken to have the
oscillator number starting from $n=0$, and thus the Lagrangian
$L_{SO(6)}$ is written in terms of modes as
\begin{align}
L_{SO(6)} = \sum^j_{m=-j} \sum^\infty_{n=0} 
V_{(mn)}^\dagger M_{(mn)} V_{(mn)} \, .
\label{so6lag}
\end{align}

From the above mode expanded Lagrangian $L_{SO(6)}$, the formal
evaluation of the path integral results in
\begin{align}
\prod^j_{m=-j} \prod^\infty_{n=0} \mathrm{Det}^{-1} M_{(mn)} \, ,
\end{align}
where $\mathrm{Det}$ involves the matrix determinant as well as the
usual functional one.  In order to get the one-loop effective action
or potential, we should first diagonalize the matrix $M_{(mn)}$.
However, the diagonalization of $M_{(mn)}$ is not an easy task,
basically due to the two constant terms $\pm 2 \sigma$ appearing in
the diagonal elements of the matrix.  Fortunately, if we consider
$M_{(mn)}$ without these two terms, it can be diagonalized without
much difficulty.  This fact naturally leads us to consider a
perturbation expansion in terms of $\sigma$.  Actually, it is not
necessary to diagonalize the matrix $M_{(mn)}$ exactly.  We are
interested in the membrane interaction in the long distance limit, and
hence the perturbation expansion is enough for our purpose.
Furthermore, since the constant parameter $\sigma$ is a small quantity
corresponding to the quantum of the area of flat membrane, it is a
good expansion parameter.

If we denote $M_{(mn)}$ without $\pm 2 \sigma$ in the diagonal
elements as $M^{(0)}_{(mn)}$, then the determinant of $M_{(mn)}$ is
written as
\begin{align}
\mathrm{Det}^{-1} M_{(mn)} =
\mathrm{Det}^{-1} M^{(0)}_{(mn)} \cdot
\det{}^{-1} \left[ 1 + 2 \epsilon \frac{E_{(n)}}{P_{(n)}} \right]
\label{detm}
\end{align}
where $\epsilon \equiv \sigma^2$ for emphasizing the parameter of
perturbative expansion,
\begin{align}
\mathrm{Det}^{-1} M^{(0)}_{(mn)} =
\det{}^{-1} \left[ -\Delta_{(n)} P_{(n)} \right] \, ,
\end{align}
and various quantities inside the functional determinants are defined
by
\begin{align}
P_{(n)} &\equiv \Delta_{(n)10} \Delta_{(n)11}
(\Delta_{(n) 1 \frac{1}{2}} + a_{n+})^2
(\Delta_{(n) 1 \frac{1}{2}} + a_{n-} )^2 \, ,
\notag \\
E_{(n)} & \equiv 
\frac{1}{3^2} \Delta_{(n)} (\Delta_{(n) 1 \frac{1}{2}} + a_{n+} )
 (\Delta_{(n) 1 \frac{1}{2}} + a_{n-} ) 
 - 2 \Delta_{(n)}^2  (\Delta_{(n) 1 \frac{1}{2}} + b_{n+} )
 (\Delta_{(n) 1 \frac{1}{2}} + b_{n-} ) \, ,
\notag \\
a_{n \pm} &\equiv -\frac{1}{6^2} \pm 
  \frac{1}{3} \sqrt{ r^2 + \sigma (2n+1)
                   +\frac{1}{3^2} \left( j+\frac{1}{2} \right)^2} \, ,
\notag \\
b_{n \pm} &\equiv -\frac{1}{6^2} \pm 
  \frac{1}{3} \sqrt{\sigma (2n+1)
                   +\frac{1}{3^2} \left( j+\frac{1}{2} \right)^2} \, .
\end{align}
This expression of the determinant is of calculable form and can be
studied perturbatively.  Then the result of path integration for
$L_{SO(6)}$ now becomes
\begin{align}
\prod^\infty_{n=0} 
\det{}^{-(2j+1)} \Delta_{(n)} \cdot \det{}^{-(2j+1)} P_{(n)}
\cdot \det{}^{-(2j+1)} 
\left[ 1 + 2 \epsilon \frac{E_{(n)}}{P_{(n)}} \right] \, .
\label{so6}
\end{align}

\subsection{Ghost sector}

The ghost sector of the quadratic action $S_2$ is described by the
Lagrangian
\begin{align}
L_G =  
&\mathrm{Tr}
\bigg[ \, 
 \dot{\bar{C}}^\dagger \dot{C} 
- (r^2 +Q^2+P^2) \bar{C}^\dagger C
- \frac{1}{3^2} \bar{C}^\dagger J^i J^i C
\notag \\
& +  \dot{\bar{C}} \dot{C}^{\dagger} 
- \bar{C} (r^2 +Q^2 +P^2) C^{\dagger}
- \frac{1}{3^2} J^i J^i  \bar{C}  C^\dagger \,
\bigg] \, . 
\end{align}
The path integration is carried out by using the same procedure taken
in the previous subsection.  If we denote the modes of the ghost
variables $C$ and $\bar{C}$ as $c_{mn}$ and $\bar{c}_{mn}$
respectively, the Lagrangian in terms of modes is obtained as
\begin{align}
L_G = 
& \sum^j_{m=-j} \sum^{\infty}_{n=0}
\left[ \,
  \dot{\bar{c}}^*_{mn} \dot{c}_{mn} 
+ \dot{\bar{c}}_{mn} \dot{c}^*_{mn} 
 - \left( r^2 + \sigma (2n+1) + \frac{1}{3^2} j(j+1) \right)
  ( \bar{c}^*_{mn} c_{mn} 
  + \bar{c}_{mn} c^*_{mn} ) \,
\right]\,.
\end{align}
The path integral for this Lagrangian is immediate, and evaluated as
\begin{align}
\prod^\infty_{n=0} \det{}^{2(2j+1)} \Delta_{(n)} \, .
\label{gdet}
\end{align}
As it should be, this ghost contribution eliminates the contributions
from unphysical gauge degrees of freedom in the results of bosonic
sector, Eqs.~(\ref{so3}) and (\ref{so6}).

\subsection{Fermionic sector}

Finally, let us consider the fermionic sector of the quadratic action.
Its Lagrangian is
\begin{align}
L_F = 
&  \mathrm{Tr} \bigg[ i \chi^\dagger \dot{\chi} 
    - \frac{i}{4} \chi^\dagger \gamma^{123} \chi
    + \frac{1}{3} \chi^\dagger \gamma^i J^i \chi
    + r \chi^\dagger (\gamma^8 \cos(t/6) + \gamma^9 \sin(t/6) ) \chi
\notag \\
& - \chi^\dagger (\gamma^4 \cos(t/6) + \gamma^6 \sin(t/6) ) \chi Q
- \chi^\dagger (\gamma^5 \cos(t/6) + \gamma^7 \sin(t/6) ) \chi P
\bigg] \, ,
\end{align}
where the matrix variable $\chi$ has been rescaled by a factor
$1/\sqrt{2}$.  Due the periodic motion of background membranes, the
Lagrangian has many trigonometric functions.  Like we have done in the
calculation of bosonic sector, we perform a transformation to the
rotating frame
\begin{align}
\chi ~ \longrightarrow ~ \Lambda \chi
\end{align}
using
\begin{align}
\Lambda = e^{-\frac{1}{12} t \gamma^{46}}e^{-\frac{1}{12} t \gamma^{57}} 
e^{-\frac{1}{12} t \gamma^{89}} \, .
\end{align}
Under this transformation, the fermionic Lagrangian becomes
\begin{align}
L_F = 
&  \mathrm{Tr} \bigg[ \, 
     i \chi^\dagger \dot{\chi}
     - \frac{i}{4} \chi^\dagger \gamma^{123} \chi
     + \frac{1}{3} \chi^\dagger \gamma^i J^i \chi
     + r \chi^\dagger \gamma^8  \chi \,
     - \chi^\dagger ( \gamma^4 \chi Q + \gamma^5 \chi P )
\notag \\
& -\frac{i}{12} \chi^\dagger 
  (\gamma^{46} + \gamma^{57} + \gamma^{89} )\chi \,
\bigg] \,.
\end{align}

In the above Lagrangian, the term $\frac{1}{3} \chi^\dagger \gamma^i
J^i \chi$ stems from the presence of the background fuzzy sphere and
should be diagonalized.  As in the case of the bosonic sector, the
diagonalization can be carried out in exactly the same way considered
in previous literatures \cite{Dasgupta:2002hx,Shin:2003np}, and thus
we will not repeat it here and just quote the result with brief
explanation. Let us first take the mode expansion of $\chi$ according
to Eq.~(\ref{decom}) as
\begin{align} 
\chi = \sum^j_{m=-j} \sum^\infty_{n=0} 
\chi_{mn} |m \rangle \langle n|
\end{align}  
The mode $\chi_{mn}$ is a complex spinor with sixteen components, and
in the representation $\mathbf{16}$ of $SO(9)$.  Under $SO(9)
\rightarrow SO(3) \times SO(6) \simeq SU(2) \times SU(4)$ reflecting
the symmetry structure of the plane wave, $\chi$ is decomposed as
$\mathbf{16} \rightarrow (\mathbf{2}, \mathbf{4}) +
(\bar{\mathbf{2}},\bar{\mathbf{4}})$.  The diagonalization acts on the
$\mathbf{2}$ and $\bar{\mathbf{2}}$ of $SU(2)$, and results in two
eigen-modes or eigen-spinors with eight independent components, say
$\pi_{mn}$ and $\eta_{mn}$, whose corresponding eigenvalues are $-j-1$
and $j$, respectively.  Here, the range of $m$ for $\pi_{mn}$
($\eta_{mn}$) is $-j \le m \le j-1$ ($-j-1 \le m \le j$).  After this
diagonalization, the Lagrangian $L_F$ becomes the sum of two
independent systems, which we call $\pi$-system, $L_\pi$, and
$\eta$-system, $L_\eta$, and is given by
\begin{align}
L_F = L_\pi + L_\eta \, ,
\end{align}  
where\footnote{In previous works
  \cite{Dasgupta:2002hx,Shin:2003np,Shin:2004az}, the eight component
  spinor notation has been used.  In this paper, we keep the sixteen
  component notation.  So, $\pi_{mn}$ and $\eta_{mn}$ are sixteen
  component spinors but have only eight independent components.}
\begin{align}
L_\pi = 
&  \sum^{j-1}_{m=-j} \mathrm{Tr} 
  \bigg[ \, 
     i \pi_m^\dagger \dot{\pi}_m
     +\frac{i}{3} \left( j+\frac{1}{4} \right) 
        \pi_m^\dagger \gamma^{123} \pi_m
     + r \pi_m^\dagger \gamma^8  \pi_m 
     - \pi_m^\dagger ( \gamma^4 \pi_m Q + \gamma^5 \pi_m P )
\notag \\
&  -\frac{i}{12} \pi_m^\dagger 
  (\gamma^{46} + \gamma^{57} + \gamma^{89} )\pi_m \,
\bigg] \,,
\notag \\
L_\eta = 
&  \sum^j_{m=-j-1} \mathrm{Tr} 
  \bigg[ \, 
     i \eta_m^\dagger \dot{\eta}_m
     - \frac{i}{3} \left( j+\frac{3}{4} \right) 
        \eta_m^\dagger \gamma^{123} \eta_m
     + r \eta_m^\dagger \gamma^8  \eta_m 
     - \eta_m^\dagger ( \gamma^4 \eta_m Q + \gamma^5 \eta_m P )
\notag \\
&  -\frac{i}{12} \eta_m^\dagger 
  (\gamma^{46} + \gamma^{57} + \gamma^{89} )\eta_m \,
\bigg] \,,
\label{pelag}
\end{align}
with the mode expansions
\begin{align}
\pi_m = \sum^\infty_{n=0} \pi_{mn} \langle n | \, , \quad
\eta_m = \sum^\infty_{n=0} \eta_{mn} \langle n | \, .
\label{fmod1}
\end{align}
For the Lagrangians $L_\pi$ and $L_\eta$, some comments are now in
order.  Firstly, in the mode expansions of $\pi_m$ and $\eta_m$, we do
not see the ket state $| m \rangle$ in the spin-$j$ representation of
$SU(2)$ anymore.  This is because the background effect due to the
fuzzy sphere has been taken into account through the diagonalization.
Secondly, there appears the term $i\pi_m^\dagger \gamma^{123} \pi_m$
in the $\pi$-system.  This is also the case in the $\eta$-system.  In
the process of calculation, this term appears originally as
$\pi_m^{(+)\dagger}\pi_m^{(+)} - \pi_m^{(-)\dagger}\pi_m^{(-)}$, where
$\pi_m^{(+)}$ ($\pi_m^{(-)}$) is the variable coming from the
diagonalization of $\mathbf{2}$ ($\bar{\mathbf{2}}$) of $(\mathbf{2},
\mathbf{4})$ ($(\bar{\mathbf{2}},\bar{\mathbf{4}})$) in the
decomposition of $\chi$.  Regarding to the action of $i\gamma^{123}$,
$\pi_m^{(\pm)}$ satisfies $i\gamma^{123}\pi_m^{(\pm)} = \pm
\pi_m^{(\pm)}$.  This simply means that $\pi_m = \pi_m^{(+)} +
\pi_m^{(-)}$ and hence we get $\pi_m^{(+)\dagger}\pi_m^{(+)} -
\pi_m^{(-)\dagger}\pi_m^{(-)} =i\pi_m^\dagger \gamma^{123} \pi_m$.

By looking at the Lagrangians $L_\pi$ and $L_\eta$ of
Eq.~(\ref{pelag}), one can easily see that they have almost the same
structure.  The $\eta$-system can be obtained from $\pi$-system by
changing the range of $m$ and replacing $j$ inside the trace by
$-j-1$.  Therefore, it is not necessary to consider the evaluation of
path integral for both of them.  From now on, we will focus on one
system, say the $\pi$-system.  The result for the $\eta$-system will
follow naturally after completing the path integral of the
$\pi$-system.

Before considering the path integral of the $\pi$-system, we would
like to note that it is convenient to change the bra vector $\langle
n|$ for the ket vector $| n \rangle$ in the mode expansion
Eq.~(\ref{fmod1}).  Such a change brings about some structural change
inside the Lagrangian.  More precisely,
\begin{align}
\mathrm{Tr} \, \pi_m^\dagger ( \gamma^4 \pi_m Q + \gamma^5 \pi_m P ) 
~ \longrightarrow ~
\pi_m^\dagger ( \gamma^4 Q - \gamma^5 P ) \pi_m
\end{align}
under 
\begin{align}
\pi_m = \sum^\infty_{n=0} \pi_{mn} \langle n | ~\longrightarrow~
\pi_m = \sum^\infty_{n=0} \pi_{mn} | n \rangle \,,
\end{align}
which can be easily checked by using Eqs.~(\ref{osc-ca}) and
(\ref{osc-alg}).  This makes the Lagrangian have more tractable form
as follows.
\begin{align}
L_\pi = 
  \sum^{j-1}_{m=-j}  \pi_m^\dagger
  \bigg[ \, 
     i \partial_t
     + \frac{i}{3} \left( j+\frac{1}{4} \right) \gamma^{123}
     + r \gamma^8  
     - \gamma^4 Q + \gamma^5 P
     -\frac{i}{12} (\gamma^{46} + \gamma^{57} + \gamma^{89} ) \,
\bigg] \pi_m \,.
\label{plag1}
\end{align}

In the above Lagrangian, there are various products of gamma matrices.
For treating them properly, we begin with the fact that $\pi_{mn}$ has
the positive chirality of $SO(9)$ because it is in $\mathbf{16}$ of
$SO(9)$, that is, $\gamma_{(9)} \pi_{mn} = \pi_{mn}$ where
$\gamma_{(9)} = \gamma^1 \gamma^2 \cdots \gamma^9$.  If we consider
the operator measuring the chirality in the $SO(6)$ symmetric space as
$\gamma_{(6)}=\gamma^4\gamma^5\gamma^6\gamma^7\gamma^8\gamma^9$, we
see that $\gamma_{(9)} = \gamma^{123} \gamma_{(6)}$.  This shows that,
for a given $SO(9)$ chirality, the eigenvalue of $\gamma^{123}$ is
automatically determined by that of $\gamma_{(6)}$, or vice versa.  In
succession, because $\gamma_{(6)} = - \gamma^{46} \gamma^{57}
\gamma^{89}$, the chiralities in 4-6, 5-7, and 8-9 planes determine
the eigenvalue of $\gamma^{123}$.  Now, let us split $\pi_m$ in terms
of the chiralities in 4-6, 5-7, and 8-9 planes as
\begin{align}
\pi_m = \sum_{s_1, s_2, s_3 = \pm} \pi_{ms_1s_2s_3}
= \sum^\infty_{n=0} \sum_{s_1, s_2, s_3 = \pm} 
\pi_{mns_1s_2s_3} | n \rangle \,,
\end{align}
where $s_1$, $s_2$, and $s_3$ represent the eigenvalues of
$\gamma^{46}$, $\gamma^{57}$, and $\gamma^{89}$, respectively.  Then,
the action of $\gamma^{46}$ on $\pi_{mns_1 s_2 s_3}$ is given by
\begin{align}
\gamma^{46} \pi_{mns_1 s_2 s_3} = i s_1 \pi_{mns_1 s_2 s_3} \, ,
\end{align}
and similarly for $\gamma^{57}$ and $\gamma^{89}$.  As for the
eigenvalue of $\gamma^{123}$, $s_1$, $s_2$, and $s_3$ determine it as
\begin{align}
\gamma^{123} = - i s_1 s_2 s_3 \, .
\end{align}

In addition to the proper handling of products of gamma matrices, the
presence of $Q$ and $P$ in the Lagrangian of Eq.~(\ref{plag1}) leads
to the mixing of modes with different oscillator number $n$.  As we
have done in the bosonic case, such mixing problem is cured by taking
an appropriate unitary transformation and then newly defined mode
expansions for some variables.  We first consider the following
unitary transformation.
\begin{align}
\zeta_{1m}^\pm &\equiv \frac{1}{\sqrt{2}} 
             (\gamma^4 \pi_{m+++}\pm i \gamma^5 \pi_{m--+}) \,,
\notag \\
\zeta_{2m}^\pm &\equiv \frac{1}{\sqrt{2}} 
             (\gamma^4 \pi_{m+--}\pm i \gamma^5 \pi_{m-+-}) \,,
\notag \\
\zeta_{3m}^\pm &\equiv \frac{1}{\sqrt{2}} 
             (\gamma^4 \pi_{m+-+}\pm i \gamma^5 \pi_{m-++}) \,,
\notag \\
\zeta_{4m}^\pm &\equiv \frac{1}{\sqrt{2}} 
             (\gamma^4 \pi_{m++-}\pm i \gamma^5 \pi_{m---}) \,.
\end{align}
These particular pairings are chosen such that the creation and
annihilation operators $a^\dagger$ and $a$ defined in
Eq.~(\ref{osc-ca}) appear independently in different terms.  After the
transformation, we find that $\zeta^\pm_{1m}$ and $\zeta^\pm_{3m}$
couple to each other as $-i \sqrt{\sigma} \zeta^{+\dagger}_{1m}
a^\dagger \gamma^5 \zeta^-_{3m} +i \sqrt{\sigma} \zeta^{-\dagger}_{1m}
a \gamma^5 \zeta^+_{3m}$ and its conjugation.  $\zeta^\pm_{2m}$ and
$\zeta^\pm_{4m}$ have the similar coupling.  Like the case of
Eq.~(\ref{mode}), the structure of couplings leads us to take the mode
expansions for $\zeta_{2m}^\pm$ and $\zeta_{3m}^\pm$ as
\begin{align}
\zeta_{2m}^\pm 
    = \sum^\infty_{n=\mp 1} \zeta_{2mn}^\pm |n \pm 1 \rangle  \,, \quad
\zeta_{3m}^\pm 
    = \sum^\infty_{n=\mp 1} \zeta_{3mn}^\pm |n \pm 1 \rangle \,,
\end{align}
while $\zeta_{1m}^\pm$ and $\zeta_{4m}^\pm$ are taken to have the
standard mode expansion.  Now, based on these mode expansions, we see
that the Lagrangian of Eq.~(\ref{plag1}) does not have any coupling
between modes with different oscillator number, and is written as
\begin{align}
L_\pi = \sum^{j-1}_{m=-j} \sum^\infty_{n=0} 
     Z_{(mn)}^\dagger F_{(mn)} Z_{(mn)} \,,
\label{pilag}
\end{align}
where $Z_{(mn)}=(\zeta^+_{1mn},\zeta^-_{1mn},\zeta^+_{2mn},
\zeta^-_{2mn},\zeta^+_{3mn},\zeta^-_{3mn},\zeta^+_{4mn},
\zeta^-_{4mn})^T$ and
\begin{align}
F_{(mn)} = 
\begin{pmatrix}
K_1 & 0 & \Gamma_n & D \\
0   & K_2 & D & \Gamma_n^\dagger \\
\Gamma_n^\dagger & D & K_3 & 0 \\
D & \Gamma_n & 0 & K_4
\end{pmatrix} \, .
\end{align}
The various quantities inside the matrix $F_{(mn)}$ are $2\times 2$
matrices and defined by
\begin{align}
& K_1 =
\begin{pmatrix}
i \partial_t + \frac{1}{3}(j+\frac{1}{2}) & \frac{1}{6} \\
\frac{1}{6} & i \partial_t + \frac{1}{3}(j+\frac{1}{2})
\end{pmatrix} \, , \quad
K_2 =
\begin{pmatrix}
i \partial_t + \frac{1}{3}j & 0\\
0 & i \partial_t + \frac{1}{3}j
\end{pmatrix} \, ,
\notag \\
& K_3 =
\begin{pmatrix}
i \partial_t - \frac{1}{3}j & 0\\
0 & i \partial_t - \frac{1}{3}j
\end{pmatrix} \, , \quad
K_4 =
\begin{pmatrix}
i \partial_t - \frac{1}{3}(j+\frac{1}{2}) & \frac{1}{6} \\
\frac{1}{6} & i \partial_t - \frac{1}{3}(j+\frac{1}{2})
\end{pmatrix} \, ,
\end{align}
and
\begin{align}
\Gamma_n =
\begin{pmatrix}
0 & -i \sqrt{2 \sigma n} \gamma^5\\
i \sqrt{2 \sigma (n+1)} \gamma^5 & 0
\end{pmatrix} \, , \quad
D =
\begin{pmatrix}
r \gamma^8 & 0\\
0 & r \gamma^8
\end{pmatrix} \, .
\end{align}  
We would like to note that, in writing the Lagrangian $L_\pi$ of
Eq.~(\ref{pilag}), we have used the reasoning similar to that leading
to $L_{SO(6)}$ of Eq.~(\ref{so6lag}), and identified symbolically
$\zeta^+_{2m-1}$ and $\zeta^+_{3m-1}$ with $\zeta^-_{2m0}$ and
$\zeta^-_{3m0}$.  So, the summation for $n$ starts from $0$.

The path integration of the $\pi$-system is now evaluated as
\begin{align}
\prod^{j-1}_{m=-j} \prod^\infty_{n=0} \mathrm{Det} F_{(mn)} \,.
\end{align}
Because of the presence of gamma matrices inside $F_{(mn)}$, the
computation of matrix determinant should be performed by using the
following matrix identity repeatedly.
\begin{align}
\begin{pmatrix} A & B\\ C & D \end{pmatrix} =
\begin{pmatrix} A & 0\\ C & 1 \end{pmatrix}
\begin{pmatrix} 1 & A^{-1} B\\ 0 & D-C A^{-1} B \end{pmatrix} \,.
\end{align}
After a bit of long computation, $\mathrm{Det} F_{(mn)}$ is obtained
as
\begin{align}
\mathrm{Det} F_{(mn)} 
&= \det \left[ Q_{(n)}
     + \frac{\epsilon}{3^2} (\partial_t^2+r^2+\frac{1}{3^2}j^2)
\right]
\notag \\
&= \det Q_{(n)} \cdot \det
\left[ 1
     + \frac{\epsilon}{3^2} 
     \frac{\partial_t^2+r^2+\frac{1}{3^2}j^2}{Q_{(n)}}
\right] \,,
\end{align}
where $\epsilon \equiv \sigma^2$ as in the bosonic case and, by using
Eq.~(\ref{kinop}), we have defined
\begin{align}
Q_{(n)} &\equiv \Delta_{(n)00} \Delta_{(n)20} 
     (\Delta_{(n)1\frac{1}{2}} + c_{n+})
     (\Delta_{(n)1\frac{1}{2}} + c_{n-}) \,,
\notag \\
c_{n \pm} &\equiv -\frac{1}{6^2} \pm 
  \frac{1}{3} \sqrt{ r^2 + \sigma (2n+1)   
+ \frac{1}{3^2} \left( j+\frac{1}{2} \right)^2 +3^2 \sigma^2} \,.
\end{align}
By using this functional determinant for a given $m$ and $n$, we can
give the result of path integration for the $\pi$-system as
\begin{align}
\prod^\infty_{n=0} \det{}^{2j} Q_{(n)} \cdot \det{}^{2j}
\left[
    1 + \frac{\epsilon}{3^2} 
    \frac{\partial_t^2+r^2+\frac{1}{3^2}j^2}{Q_{(n)}}
\right] \,.
\label{pathp}
\end{align}

Finally, we consider the path integration of the $\eta$-system.  As
mentioned earlier, the $\eta$-system is the same with the $\pi$-system
if we change the range of $m$ and take the replacement $j \rightarrow
-j-1$.  This means that we can get the result of path integral for the
$\eta$-system without any further calculation.  Then, from the result
of $\pi$-system, Eq.~(\ref{pathp}), we see that the path integration
of the $\eta$-system leads to
\begin{align}
\prod^\infty_{n=0} \det{}^{2j+2} \tilde{Q}_{(n)} \cdot \det{}^{2j+2}
\left[
  1 + \frac{\epsilon}{3^2} 
  \frac{\partial_t^2+r^2+\frac{1}{3^2}(j+1)^2}{\tilde{Q}_{(n)}}
\right] \,,
\label{pathe}
\end{align}
where
\begin{align}
\tilde{Q}_{(n)} \equiv
\Delta_{(n)01} \Delta_{(n)21} (\Delta_{(n)1\frac{1}{2}} + c_{n+})
(\Delta_{(n)1\frac{1}{2}} + c_{n-}) \,.
\end{align}

\section{Effective potential}
\label{epotential}

We have evaluated the path integral for the bosonic, ghost, and
fermionic sectors in the last section, and obtained the functional
determinants given in Eqs.~(\ref{so3}), (\ref{so6}), (\ref{gdet}),
(\ref{pathp}), and (\ref{pathe}).  The multiplication of them now
gives $\exp (i \Gamma^{\text{1-loop}}_{\mathrm{eff}})$, where
$\Gamma^{\text{1-loop}}_{\mathrm{eff}}$ is the one-loop effective
action describing the interaction between the fuzzy sphere and flat
membranes.  In this section, we obtain the one-loop effective
potential $V_{\mathrm{eff}}$ from the effective action via the
relation $\Gamma^{\text{1-loop}}_{\mathrm{eff}} = - \int dt
V_{\mathrm{eff}}$.

As we have seen in the last section, some functional determinants
obtained after the formal path integral are not of fully factorized
form.  Although it is so, they can be studied perturbatively in terms
of the small parameter $\epsilon$ which is defined by $\sigma^2$.  By
the way, the structure of functional determinants containing
$\epsilon$ tells us that the $\epsilon$ expansion is nothing but the
large distance expansion.  This matches precisely with our purpose,
because our prime interest is the leading order effective potential in
the large distance limit.  Here we would like to note that the large
distance means large $r$ compared to the size $N_1$ of the fuzzy
sphere, that is, $r \gg N_1$.

At this point, apart from the numerical factor, one may actually guess
the form of the leading order potential for the background
configuration considered here.  The guess is that the potential is
attractive and behaves as $1/r^5$ at the leading order.  However, as
we will see, the calculation leads to an unexpected result that the
leading order behavior is not $1/r^5$ but $1/r^3$.

Then we first consider the effective potential at the lowest order in
$\epsilon$.  From the functional determinants, we can obtain the
following potential without much difficulty.
\begin{align}
&\sum^\infty_{n=0}
\bigg[ 
4j \sqrt{m^2_{10}} + 2(2j+2) \sqrt{m^2_{11}}
+2(2j+1) 
   \sqrt{m^2_{1\frac{1}{2}}
         +\frac{1}{6^2} + 
         \frac{1}{3}\sqrt{m^2_{1 \frac{1}{2}}}
        }
\notag \\
& +2(2j+1) 
   \sqrt{m^2_{1\frac{1}{2}}
         +\frac{1}{6^2} - 
         \frac{1}{3}\sqrt{m^2_{1\frac{1}{2}}}
        }
- 2j \sqrt{m^2_{00}} - 2j \sqrt{m^2_{20}}
\notag \\
&- (2j+2) \sqrt{m^2_{01}} - (2j+2) \sqrt{m^2_{21}}
-2(2j+1) \sqrt{m^2_{1\frac{1}{2}} + \frac{1}{6^2} +
 \frac{1}{3}\sqrt{m^2_{1\frac{1}{2}} + 3^2 \sigma^2}}
\notag \\
& -2(2j+1) \sqrt{m^2_{1\frac{1}{2}} + \frac{1}{6^2} -
 \frac{1}{3}\sqrt{m^2_{1\frac{1}{2}} + 3^2 \sigma^2}} \,
\bigg] \,,
\end{align}
where
\begin{align}
m^2_{\alpha\beta} \equiv 
  r^2 + \sigma (2n+\alpha) +\frac{1}{3^2}(j+\beta)^2 \,.
\end{align}
The potential is expressed as an infinite sum over $n$.  This may
cause to worry about convergence.  However, if we investigate the
potential at large $n$, we find that it behaves as $n^{-3/2}$ and thus
the summation is well-defined.  The sum over $n$ can be performed by
adopting the Euler-Maclaurin formula
\begin{align}
\sum^\infty_{n=0} f(n) = \int^\infty_0 dx f(x) + \frac{1}{2} f(0)
-\frac{1}{12} f'(0) + \frac{1}{720}f'''(0) + \dots
\label{emf}
\end{align}
which is valid when $f$ and its derivatives vanish at infinity.  After
the summation, if we expand the resulting potential in terms of large
$r$, we obtain
\begin{align}
\frac{N_1 \sigma}{r} 
- \frac{1}{432}(24 j^2 + 24 j + 13) \frac{N_1 \sigma}{r^3}
+ \mathcal{O} \left( \frac{1}{r^5}\right) \,,
\label{pot0}
\end{align}
where $N_1 = 2j+1$ has been used.

We turn to the effective potential at the first order in $\epsilon$.
Let us first consider the contribution from the bosonic part, that is,
from Eq.~(\ref{so6}).  From the relation
\begin{align}
\det{}^a (1+ A) = \exp [ \, a \mathrm{tr} \ln (1+A) \, ]
= \exp[ \, a \mathrm{tr} A - a \mathrm{tr} A^2/2 + \dots ]
\end{align}
where $\mathrm{tr}$ is the functional trace, we see that the relevant
contribution to the effective potential is $-2i N_1 \sigma^2
\sum^\infty_{n=0} \mathrm{tr} E_{(n)}/P_{(n)}$.  The trace calculation
of this is transformed to an integration in momentum space.  After
evaluating the integration, the Euler-Maclaurin formula (\ref{emf})
and the expansion in terms of large $r$ then lead us to have the
following bosonic contribution to the effective potential at
$\epsilon^1$-order.
\begin{align}
-2i N_1 \sigma^2 \sum^\infty_{n=0} \int^\infty_{-\infty} 
\frac{d \omega}{2\pi} \frac{E_{(n)}}{P_{(n)}} =
-\frac{N_1 \sigma}{r}
+ \frac{1}{216}(12j^2+12j+1) \frac{N_1 \sigma}{r^3}
+ \mathcal{O} \left( \frac{1}{r^5}\right) \,,
\label{pot-bos}
\end{align}
where $\omega$ is the conjugate momentum of time $t$, and
$\partial_t^2$ inside $E_{(n)}/P_{(n)}$ is understood to be replaced
by $-\omega^2$.  This contribution shows explicitly that the $1/r$
term on the right hand side exactly cancels that of the lowest order
potential (\ref{pot0}).  Thus, up to this point, the leading order
interaction for large $r$ is of $1/r^3$ type.

Another contribution at the first order in $\epsilon$ comes from
fermionic part given by Eqs.~(\ref{pathp}) and (\ref{pathe}).  If we
follow the same steps taken in the previous paragraph, we get the
contributions from the $\pi$-system (\ref{pathp}) as
\begin{align}
-\frac{1}{108}\frac{j \sigma}{r^3}
+ \mathcal{O} \left( \frac{1}{r^5}\right) \,,
\end{align}
and from the $\eta$-system as
\begin{align}
-\frac{1}{108}\frac{(j+1) \sigma}{r^3}
+ \mathcal{O} \left( \frac{1}{r^5}\right) \,.
\end{align}
Thus, the total contribution from the fermionic part to the 
effective potential is
\begin{align}
-\frac{1}{108}\frac{N_1 \sigma}{r^3}
+ \mathcal{O} \left( \frac{1}{r^5}\right) \,.
\label{pot-fer}
\end{align}

If we gather the results obtained up to now, Eqs.~(\ref{pot0}),
(\ref{pot-bos}), and (\ref{pot-fer}), then we see that the one-loop
effective potential in the large distance limit becomes
\begin{align}
V_{\mathrm{eff}} = - \frac{5}{144} \frac{N_1 \sigma}{r^3} + 
\mathcal{O} \left( \frac{1}{r^5}\right) \,.
\label{effpot}
\end{align}
This is the effective potential up to the first order in $\epsilon$.
Here, one may wonder if the contributions coming from higher
$\epsilon$ order correct the numerical factor of the leading order
term or make $r^{-5}$ the leading interaction term for large $r$ by
canceling the $r^{-3}$ term in (\ref{effpot}).  However, if we
contemplate Eqs.~(\ref{so6}), (\ref{pathp}), and (\ref{pathe}) and
perform a simple power counting, it is not difficult to see that the
higher $\epsilon$ order leads to at most the interaction of
$\mathcal{O} (r^{-5})$.  Therefore, the leading $r^{-3}$ type
interaction of (\ref{effpot}) remains intact even if we consider the
contributions from higher $\epsilon$ order, and it is one-loop exact.

The one-loop effective potential (\ref{effpot}) shows that there is an
interaction between the fuzzy sphere and flat membranes, which is
attractive.  At this point, let us recall the background
configuration, (\ref{fuzzy}) and (\ref{flat}).  Although it is taken
such that the fuzzy sphere membrane moves around the flat one, it is
basically a `static' one in a sense that the distance $r$ between two
membranes does not change as time goes by.  The presence of an
attractive interaction in this `static' configuration strongly
suggests that our membrane configuration is quite similar to the usual
D2-D0 system where two D-branes are apart with a distance $r$.  Since
D0-brane is simply a graviton from the eleven-dimensional viewpoint,
what we can conclude from this similarity is that the fuzzy sphere
membrane behaves like a graviton, that is, a giant graviton.  Thus the
present calculation gives one more check about the interpretation of
the fuzzy sphere membrane as a giant graviton.

One interesting fact is that the leading order interaction at 
large distance is of $r^{-3}$ type rather than $r^{-5}$ type.
Usually, the increase of $r$ power is related to the delocalization
or smearing of brane in some directions.  As for the present case,
the $r$ power increases by two from the expected power.  This
implies that one of two membranes is delocalized in two spatial
directions.  From the background configuration, it is not so difficult
to guess that the flat membrane corresponds to such delocalized
brane.  The flat membrane of Eq.~(\ref{flat}) is taken to span and 
spin in four dimensions.  So two extra directions are required for 
its description.  We interpret that this brings about the
delocalization or smearing effect which manifests in the interaction
potential.

\section{Conclusion and discussion}
\label{finalsec}

We have studied the interaction between flat and fuzzy sphere
membranes in plane-wave matrix model and computed the one-loop
effective potential at large distance limit.  Similar to the usual
D2-D0 system or more directly the membrane-graviton system in eleven
dimensions \cite{Aharony:1996bh}, the interaction is non-vanishing and
attractive.  This shows that the fuzzy sphere membrane behaves like a
graviton, the giant graviton.  So, our result gives one more evidence
about the interpretation of fuzzy sphere membrane as a giant graviton.
By the way, interestingly enough, the leading interaction at large
distance $r$ is not the expected $r^{-5}$ but $r^{-3}$ type.  We have
interpreted this type of interaction as that incorporating the
delocalization or smearing effect due to the configuration of the flat
membrane which spans and spins in four dimensional space.

In fact, the smearing effect has been already reported in the
supergravity side \cite{Bain:2002nq,Mas:2003uk}.  In the plane-wave
background, it has been observed that some supergravity solutions show
the delocalization or smearing of branes in some directions.  Our
result may be the first explicit realization of the smearing effect in
the matrix model side.

The effective potential we have obtained gives an attractive
interaction.  So, it is natural to expect that the final configuration
may be the bound state of the flat and fuzzy sphere membranes.
Although our effective potential is valid only at large distance and
we do not know what happens at small distance, the bound state is
quite interesting if it is possible.  As for the D2-D0 system, two
D-branes form a bound state at the final stage and D0-brane is
realized as the magnetic field on the worldvolume of D2-brane.
Contrary to the D0-brane, the fuzzy sphere membrane is not point-like
and has a size.  If it is really bound to the flat membrane, it is
very interesting to ask about the fate of two membranes.  At present,
this is an open question.  We hope to return to this issue in a near
future.

\section*{Acknowledgments}

The author would like to thank the theory group at KEK
for worm hospitality.  This work was supported by the Korea Science
and Engineering Foundation (KOSEF) grant funded by the Korea
government (MEST) through the Center for Quantum Spacetime (CQUeST) of
Sogang University with grant number R11-2005-021.
This work was also supported by the Korea Science and
Engineering Foundation (KOSEF) grant funded by the Korea
government (MEST), No.~R01-2008-000-21026-0.

\end{document}